\documentclass[reprint,floatfix,superscriptaddress,noshowpacs]{revtex4-1}
\usepackage{graphics,graphicx,epsfig,amsmath,amssymb,epstopdf,wasysym,comment}
\usepackage{hyperref}
\usepackage[dvipsnames]{xcolor}
\allowdisplaybreaks

\newcommand{\be}{\begin{equation}}
\newcommand{\ee}{\end{equation}}
\newcommand{\bea}{\begin{eqnarray}}
\newcommand{\eea}{\end{eqnarray}}
\newcommand{\ba}{\begin{eqnarray*}}
\newcommand{\ea}{\end{eqnarray*}}
\newcommand{\dagga}{{\phantom{\dagger}}}

\newcommand{\bk}{\mathbf{k}}

\newcommand{\br}{\mathbf{r}}

\newcommand{\dis}{\displaystyle}

\newcommand{\fract}[2]{\frac{\dis #1}{\dis #2}}

\newcommand{\eqn}[1]{(\ref{#1})}

\newcommand{\ep}{\epsilon}

\newenvironment{eqs}%
{\begin{equation} \begin{aligned}}%
{\end{aligned} \end{equation} }
\newcommand{\beal}{\begin{eqs}}
\newcommand{\eal}{\end{eqs}}
\newcommand{\bw}{\begin{widetext}}
\newcommand{\ew}{\end{widetext}}

\newcommand{\bd}[1]{\boldsymbol{#1}}
\newcommand{\ext}{{\text{ext}}}
\newcommand{\sys}{{\text{sys}}}

\begin{document}
\title{
  The perils of minimal coupling to electromagnetic field in quantum many-body systems
}

\author{Jan Skolimowski}
\affiliation{International School for
  Advanced Studies (SISSA), Via Bonomea
  265, I-34136 Trieste, Italy} 
\author{Adriano Amaricci}
\affiliation{CNR-IOM DEMOCRITOS, Istituto Officina dei Materiali,
Consiglio Nazionale delle Ricerche, Via Bonomea 265, I-34136 Trieste, Italy}
\affiliation{International School for
  Advanced Studies (SISSA), Via Bonomea
  265, I-34136 Trieste, Italy} 
\author{Michele Fabrizio}
\affiliation{International School for
  Advanced Studies (SISSA), Via Bonomea
  265, I-34136 Trieste, Italy} 

\begin{abstract}
  Consistency with the Maxwell equations determines how matter must be
  coupled to the electromagnetic field (EMF) within the minimal coupling
  scheme.  Specifically, if the Hamiltonian includes just a short-range
  repulsion among the conduction electrons,  
  as is commonly the case for models of correlated metals,
  those electrons must be coupled to the full internal EMF, 
  whose longitudinal and transverse components are self-consistently related 
  to the electron charge and current densities through Gauss's and
  circuital laws, respectively.
  Since such self-consistency relation is
  hard to implement  when modelling the non-equilibrium dynamics caused
  by the EMF, as in pump-probe experiments, it is common to replace in
  model calculations the internal EMF by the external one. Here we show
  that such replacement may be extremely dangerous, especially when the
  frequency of the external EMF is below the intra-band plasma edge.
\end{abstract}
                         
\maketitle

\section{Introduction}
Modern ultrafast time resolved pump-probe spectroscopy offers the
possibility to access the real-time dynamics of a material perturbed
by a  laser pulse, thus providing information complementary to
more traditional  experimental techniques. Furthermore, properly tailoring 
the pump pulse allows ultrafast photoinducing phase transitions into
states that may not even exist in thermal equilibrium
\cite{Hidden-1,Dragan-Science2014}.
Strongly correlated materials appeared as ideal candidates for such
kind of
experiments~\cite{Iwai2003PRL,Mansart2010EEL,Lantz2017NC,Giannetti2016AIP,Basov_2017,Lantz2017NC,Cavalleri_2018,Braun2018NJOP,Singer2018PRL,Giorgianni2019NC}, 
because of their rich phase diagrams that include different insulating
and  conducting states, often displaying notable properties, such as
high-T$_c$
superconductivity~\cite{Fausti2011S,Dal-Conte2012S,PhotoSC_Cavalleri}.

The experimental activity has, in turn, stimulated a great theoretical 
effort aimed to interpret the measurements, as well as to
achieve control over new states of matter that might be stabilised by a
properly designed laser pulse, see for instance
Refs.~\cite{Tsuji2008PRB,Tsuji-PRL2011,Martin-RMP2014,Oka-Floquet-2019}. 
Evidently, this task requires a proper treatment of the interaction with
the electromagnetic field. 
The minimal coupling scheme, describing the 
light-matter interaction when only the monopole of the charged
particles is taken into account, is explicitly derived in many
textbooks~\cite{Schiff,Mahan} and routinely used to model the 
electromagnetic field coupling in electronic systems.
However, its precise meaning in the case of many-body systems is often
largely overlooked, ultimately leading to a possible fallacious
description of the effects of light. 

In the following we shall show that the inconsiderate use of the
simple minimal coupling recipe hides in reality some approximations
which are not always justified. In particular we review a correct
treatment of the electromagnetic field coupling in a system of electrons within
linear regime, pointing
out the implicit assumptions which may not be verified in metallic
systems. We discuss a simple paradigmatic, yet generic, case in which
the difference in the treatment of the external field can lead to
rather different results.    

\section{Discussion}
We assume a system of charged particles in presence of external sources of the electromagnetic field that can be described in terms of
the external scalar, $\phi_\ext(\br,t)$, and vector potentials, $\bd{A}_\ext(\br,t)$. We decompose $\bd{A}_\ext(\br,t)=\bd{A}_{||\ext}(\br,t) +
\bd{A}_{\perp\ext}(\br,t)$, where $\bd{A}_{||\ext}(\br,t)$ and $\bd{A}_{\perp\ext}(\br,t)$ are the longitudinal and transverse components, respectively. In the following we shall work in the Coulomb gauge $\bd{\nabla}\cdot\bd{A}_\ext(\br,t)=0$, so that the vector potential is purely
transverse~\cite{Mahan}.

Since our system is made of charged particles, they actually feel ``internal'' scalar and vector potentials,
 $\phi(\br,t)$ and $\bd{A}_\perp(\br,t)$, respectively, which do not in general coincide with the external ones. 
Because of the linearity of the Maxwell equations, we can 
express such internal fields as: 
\beal
\phi(\br,t) &= \phi_\ext(\br,t) + \phi_\sys(\br,t)\,,\\
\bd{A}_\perp(\br,t) &= \bd{A}_{\perp\ext}(\br,t) + \bd{A}_{\perp\sys}(\br,t)\,,
\label{int.vs.ext}
\eal
where the system $\phi_\sys(\br,t)$ and 
$\bd{A}_{\perp\sys}(\br,t)$ potentials are obtained through the Gauss's law 
\beal
-\nabla^2\phi_\sys(\br,t) &= 4\pi\,\rho_\sys(\br,t)\,,\label{Gauss}
\eal
and the circuital law
\beal
\bigg(\fract{\partial^2}{\partial t^2} - c^2\,\nabla^2\bigg)\,\bd{A}_{\perp\text{sys}}(\br,t) &= 
4\pi\,c\,\bd{J}_{\perp\text{sys}}(\br,t) \,,\label{self-circuit}
\eal
with $\rho_\sys(\br,t)$ the system charge density, and 
$\bd{J}_{\perp\text{sys}}(\br,t)$ the transverse component of the system 
current density.
Consequently, the internal gauge-invariant electric and magnetic fields 
are defined in terms of the internal scalar and vector potentials through 
\beal
\bd{E}_{||}(\br,t) &= 
- \bd{\nabla}\phi(\br,t)\,,\\
\bd{E}_{\perp}(\br,t) &= 
-\frac{1}{c}\,\fract{\partial \bd{A}_{\perp}(\br,t)}{\partial t}\,,\\
\bd{B}_\perp(\br,t) &= \bd{\nabla}\wedge\bd{A}_\perp(\br,t)\,.
\eal 

Using the above definitions, the Hamiltonian that describes our 
system coupled to the electromagnetic field,
which we assume to be classical, 
reads, in the minimal coupling scheme~\cite{Schiff,Mahan} and neglecting the Zeeman term,   
\bw
\beal
\mathcal{H} &= \int d\br \,\Bigg\{ \Bigg[\sum_\sigma\,
\Psi^\dagger_\sigma(\br)\,\fract{1}{2m}\bigg(-i\hbar\,\bd{\nabla} + \fract{e}{c}\,\bd{A}(\br,t)\bigg)^2\,
\Psi^\dagga_\sigma(\br)\,\Bigg] + V(\br)\,\Psi^\dagger_\sigma(\br)\,\Psi^\dagga_\sigma(\br)
 \,\Bigg\}\\
&\qquad + \fract{e^2}{2}\,\sum_{\sigma\sigma'}\int d\br\,d\br'\, 
\Psi_\sigma^\dagger(\br)\,\Psi_{\sigma'}^\dagger(\br')\;\fract{1}{\;\big|\br-\br'\big|\;}\;
\Psi_{\sigma'}^\dagga(\br')\,\Psi_\sigma^\dagga(\br)
+\int d\br\,\phi_\ext(\br,t)\,\rho(\br,t) 
\,,\label{Ham}
\eal
\ew
where $\Psi^\dagga_\sigma(\br)$ is the Fermi field of spin $\sigma$ electrons, 
$V(\br)$ the periodic potential of an underlying lattice of immobile
ions that also provide a
positive charge density, $\rho_\text{ion}(\br)$, neutralising 
the electron one. Thus we have:  
$\rho(\br)\equiv \rho_\text{ion}(\br) -e\,\sum_\sigma
\Psi_\sigma^\dagger(\br)\,\Psi_\sigma^\dagga(\br)$. 

It is worth emphasising that Eqs.~\eqn{Gauss} and \eqn{self-circuit}, where 
\beal
\rho_\sys(\br,t) &= \langle\,\rho(\br)\,\rangle\,,\\
\bd{J}_{\perp\text{sys}}(\br,t) &= -c\,\big\langle\;\fract{\delta \mathcal{H}}{\delta \bd{A}_\perp(\br,t)}\;
\big\rangle\,,
\eal
to be verified require that 
\begin{enumerate}
\item one must explicitly include the Coulomb interaction among the electrons in order for the Hamiltonian \eqn{Ham} to involve 
only the external longitudinal field $\phi_\ext(\br,t)$;
\item the transverse vector potential $\bd{A}_\perp(\br,t)$ is the
  internal one, i.e., the sum of the external potential
  $\bd{A}_{\perp\text{ext}}(\br,t)$ plus  the one generated by the
  electrons, $\bd{A}_{\perp\text{sys}}(\br,t)$,  through Eq.~\eqn{self-circuit}.  
\end{enumerate}
The issue is that both points {1.} and {2.} make it difficult
modelling the system dynamics during and after the action of an
electromagnetic pulse. To proceed further, some approximations have to
be assumed. 
Concerning point {1.}, we note that correlated materials are commonly
described in terms of lattice models with short range
electron-electron interactions, e.g., the paradigmatic Hubbard model.
Although such models are in general not exactly solvable, powerful
techniques are available to investigate them in controlled
approximation schemes, such as dynamical 
mean field theory (DMFT)\cite{DMFT}, originally designed to treat
just short range interaction.
Several attempts to add non-local correlations in equilibrium DMFT
have been put forward~\cite{Biermann2014,Casula2016,Toschi-RMP2018},   
still the inclusion of the true long-range Coulomb interaction remains a
serious challenge. The extension of some of those attempts 
to the out-of-equilibrium regime has been achieved in simple cases\cite{Martin-RMP2014,Golez2017PRL,Golez2019PRB},
but a more systematic development and a proper description of the
dynamics in presence of a longitudinal field is yet to come. 

However, since the laser frequency in experiments usually ranges from
far to near infrared, i.e. wavelengths $\lambda \geq 1\mu\text{m}$, the difference between longitudinal and transverse components of the electromagnetic field is negligible. In this case, one can in principle focus only on the transverse response, 
which is seemingly less sensitive to the long range tail of the
Coulomb repulsion \cite{irreducible}.

However, the long range nature of the coupling to the transverse field 
is hidden in point {2.} above, which entails 
the self-consistency condition \eqn{self-circuit} that is not easy to
implement in an actual calculation.
One can avoid that self-consistency by treating the transverse 
field quantum mechanically, and integrating out the photons. The result 
would be that only the external vector potential would now appear in
the minimal coupling scheme,
at the cost of introducing a current-current interaction among the
electrons, non-local both in time and space.
At the end, one faces again the same problems as in the longitudinal
response, worsened by the non-locality in time.

In view of the above difficulties, it is rather common to simply ignore points 
{1.} and {2.} above, and just consider models of correlated electrons
interacting via a short range repulsion, and minimally coupled to a
uniform vector potential assumed to coincide with
the external one, $\bd{A}(t)= \bd{A}_\text{ext}(t)$, see, for example, 
Refs.~\cite{Martin-RMP2014} and \cite{Oka-Floquet-2019}.

Our aim here is not to revise all results that have been so far
obtained under those simplifications,
but just to select few examples that can be explicitly worked out and
where the difference between taking or not into account points {1.}
and {2.} is most dramatic.

For simplicity, we consider the half-filled single-band Hubbard model 
in a three dimensional cubic lattice with nearest neighbour hopping 
$-\mathsf{t}$, and 
in the presence of a uniform AC vector potential.
Using the Peierls substitution method, the Hamiltonian reads  
\beal
\mathcal{H}(t) &= \sum_{\bk\sigma}\,\varepsilon\Big(\bk+\fract{e}{\hbar c}\,\bd{A}(t)\Big)\,c^\dagger_{\bk\sigma}\,c^\dagga_{\bk\sigma}
\\
&\qquad + \fract{1}{2}\,\sum_{i,j}\,\big(n_i-1\big)\,U_{ij}\,\big(n_j-1\big)
\\
&= \mathcal{H}_0 
+ \sum_{\bk\sigma}\,\bigg(\varepsilon\Big(\bk+\fract{e}{\hbar c}\,\bd{A}(t)\Big)-\ep(\bk)\bigg)\,c^\dagger_{\bk\sigma}\,c^\dagga_{\bk\sigma}\\
&= \mathcal{H}_0 + \delta\mathcal{H}(t)\,,\label{Ham-Hubb}
\eal
where 
\beal
\varepsilon(\bk) &= -2\mathsf{t}\,\sum_{n=1}^3\,\cos k_n a\,,
\eal
with $a$ the lattice constant, $\bd{A}(t)$ the internal vector
potential, transverse and longitudinal loosing their meaning in the
present uniform case, and $U_{ij}$ the Coulomb interaction.  

Focusing on the response to the internal $\bd{A}(t)$, we can sensibly discard
the long-range tail of $U_{ij}$ \cite{irreducible}, and thus  
approximate $U_{ij}=U$ if $i=j$, and zero otherwise, i.e., the
standard local Hubbard repulsion.

We assume to be in a linear response regime, and that the probing measurement 
is performed well beyond the characteristic
relaxation time of the system~\cite{Perfetti2008NJOP,Novelli2014NC,Wall2018S}.
With those assumptions the Hamiltonian is 
\bw
\beal
\mathcal{H}(t) &\simeq \sum_{\bk\sigma}\,\Bigg(\varepsilon(\bk) 
\bigg(1- \fract{e^2 a^2}{\,6\hbar^2 c^2\,}\;\bd{A}(t)\cdot\bd{A}(t)\bigg)
+ \fract{e}{\hbar c}\;\fract{\partial\varepsilon(\bk)}{\partial \bk}\cdot \bd{A}(t)
\Bigg)\,c^\dagger_{\bk\sigma}\,c^\dagga_{\bk\sigma}
 + \fract{U}{2}\,\sum_{i}\,\big(n_i-1\big)^2\,,\label{Ham-Hubb-1}
\eal
\ew
and the equation relating the internal field to the external one has the simple solution, in the frequency space, 
\beal
\bd{A}(\omega) &= \fract{\bd{A}_\ext(\omega)}{\ep(\omega)}\;,
\label{Aint.vs.Aext}
\eal 
with the uniform dielectric constant 
\beal
\epsilon(\omega) &= 1 + i\,\fract{4\pi}{\omega}\;\sigma(\omega)\,,
\label{dielectric}
\eal
where $\sigma(\omega)$ is the optical conductivity that, 
in linear response, is defined by
\beal
\bd{J}_\sys(\omega) &= \sigma(\omega)\,\bd{E}(\omega) =
\fract{i\omega}{c}\;\sigma(\omega)\,\bd{A}(\omega)\,,   
\label{optical}
\eal
and can be calculated through the current-current response function. 
We shall here focus on two physical quantities that can be readily
obtained once the optical conductivity and the dielectric constant are
known.

\begin{figure}
\centerline{\includegraphics[width=0.5\textwidth]{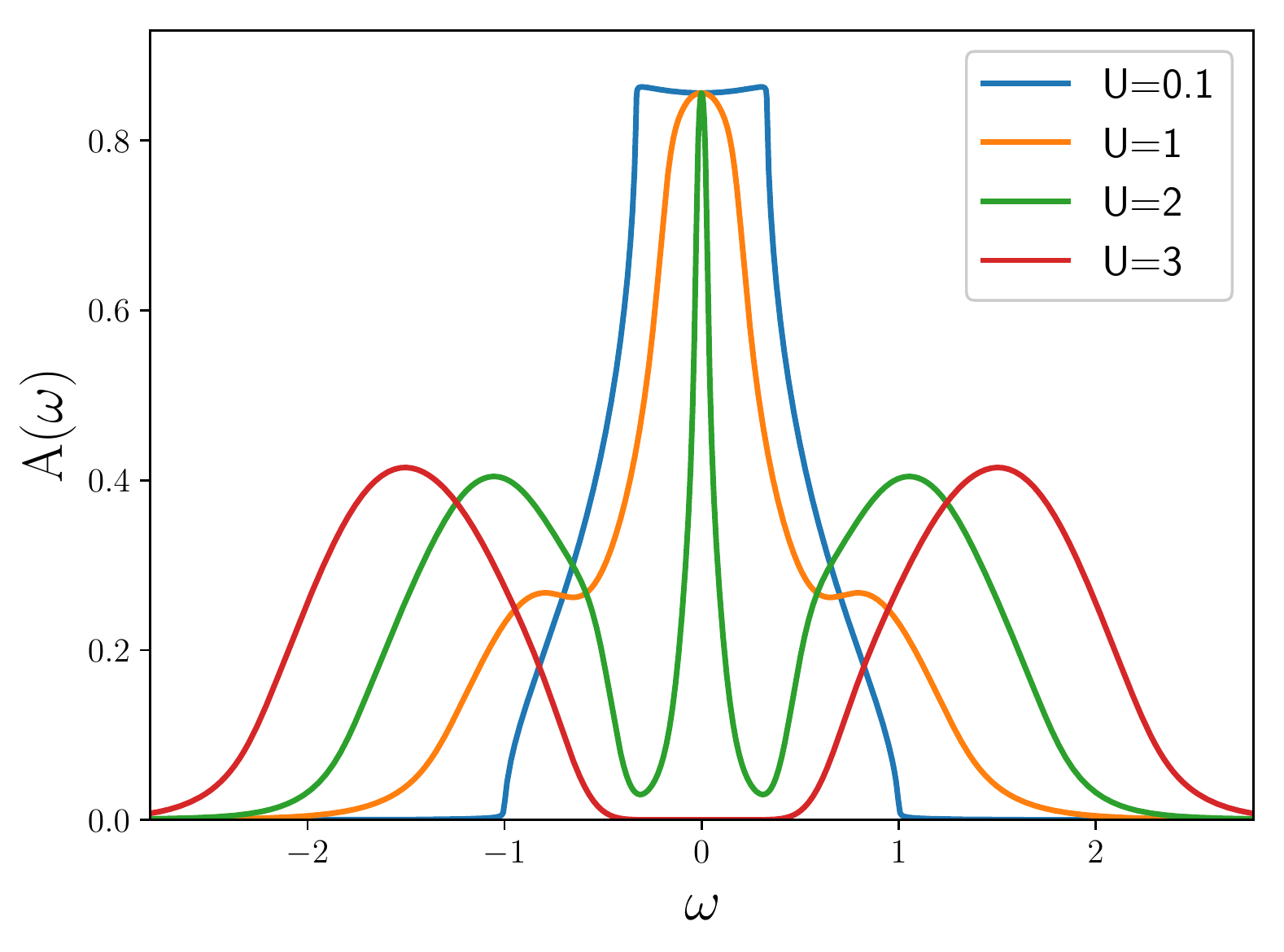}}
\caption{(Color online) Evolution of the local spectral function
  $A(\omega)$ of the single-band Hubbard model on the cubic
  lattice. Data are for different values of $U=0.1$, $1$ ,$2$, $3$,
  from the weakly correlated metal up to the Mott insulator.}
\label{LDOS}
\end{figure}

The first is the expectation value of the hopping 
\beal
T\big(\bd{A}(t)\big) &= \big\langle\;
\fract{\partial \mathcal{H}(t)}{\partial \mathsf{t}}\;\big\rangle
\\
&\simeq T(0)\,\bigg(1-\fract{e^2 a^2}{\,6\hbar^2 c^2\,}\;\bd{A}(t)\cdot\bd{A}(t)\bigg)\,,
\eal
which is renormalised downwards by the electromagnetic field, with 
potentially interesting consequences, see, e.g., \cite{Tsuji-PRL2011,Tsuji2008PRB,Ono-PRB2017}. 
We choose to quantify this reduction through the relative variation of
the hopping expectation value averaged over one period
$\tau=2\pi/\omega$ of a monochromatic field of frequency $\omega$,
which reads
\bea
\fract{\delta T}{T}&=&\int_0^{\tau} \fract{dt}{\tau}\; 
\left|\fract{\;T\big(\bd{A}(t)\big)-T(0)\;}{T(0)}\right| 
= \fract{e^2 a^2}{\,12\hbar^2 \omega^2\,}\;
\big|\bd{E}(\omega)\big|^2\nonumber \\
&=&  \fract{e^2 a^2}{\,12\hbar^2 \omega^2\,}\;
\fract{\;\big|\bd{E}_\ext(\omega)\big|^2\;}{\big|\ep(\omega)\big|^2}\;.
\eea
The reduction thus becomes significant when $\delta T/T =1$, 
which corresponds to a threshold field  
\beal
\big|\bd{E}^\text{th}_\ext(\omega)\big| = \fract{\sqrt{12}\,\hbar\omega}{ea}\,\big|\ep(\omega)\big|
\,.\label{uno}
\eal
We observe that, if one discards point \textbf{2.}, i.e., assumes 
$\bd{A}(t)$ to coincide with $\bd{A}_\ext(t)$, the threshold field changes 
into $\big|\bd{E}^\text{th}_\ext(\omega)\big|_\text{appx}$ related to the true one of Eq.~\eqn{uno} through 
\beal
\fract{\;\big|\bd{E}^\text{th}_\ext(\omega)\big|_\text{appx}\;}{\big|\bd{E}^\text{th}_\ext(\omega)\big|}  = \fract{1}{\;\big|\ep(\omega)\big|\;}
\equiv Y_1(\omega)
\,.
\label{uno-wrong}
\eal
We shall use $Y_1(\omega)$ as first estimate of the error one can do by 
replacing the internal vector potential with the external one in the minimal 
coupling scheme \eqn{Ham}. 

The other physical quantity we consider is the power dissipated by 
the monochromatic electromagnetic field during one period, defined as, see
Eq.~\eqn{Ham-Hubb},  
\beal
P(\omega) &= \int_0^\tau \fract{dt}{\tau}\; \fract{\partial}{\partial t}\,
\langle\;\mathcal{H}_0\;\rangle 
= -i\,\int_0^\tau \fract{dt}{\tau}\; \langle\;\Big[\mathcal{H}_0\,,\,
\mathcal{H}(t)\Big]\;\rangle\\
&= \fract{1}{2}\,\text{Re}\,\sigma(\omega)\;\big|\bd{E}(\omega)\big|^2
= \fract{1}{2}\,\fract{\;\text{Re}\,\sigma(\omega)\;}{\;\big|\ep(\omega)\big|^2\;}\;\big|\bd{E}_\ext(\omega)\big|^2\,.
\label{due}
\eal

As before, if one uses $\bd{A}_\ext(t)$ instead of $\bd{A}(t)$ in the 
Hamiltonian, the power dissipated takes the approximate expression 
\beal
P_\text{appx}(\omega) &= \fract{1}{2}\,\text{Re}\,\sigma(\omega)\;\big|\bd{E}_\ext(\omega)\big|^2\,.\label{due-wrong}
\eal
$P(\omega)$ is the energy of the electromagnetic field that is actually absorbed by the system per unit time. If the system thermalises, such supplied energy is transformed into heat that yields an effective temperature raise $\Delta T$ given by 
\beal
\Delta T &= \fract{P(\omega)\,\tau_\text{pulse}}{c_V}\;,
\eal
where $\tau_\text{pulse}$ is the laser pulse duration, and $c_V$ the system specific heat. Seemingly, if one identifies the vector potential in the minimal coupling with the external one, and thus uses the approximate expression \eqn{due-wrong}, the temperature raise changes into $\Delta T_\text{appx}$, where 
\beal
\fract{ \Delta T_\text{appx}}{\Delta T} &= 
\fract{P_\text{appx}(\omega)}{P(\omega)} = \big|\ep(\omega)\big|^2
\equiv Y_2(\omega)
\,.
\label{Delta T vs.}
\eal
$Y_2(\omega)$ is the other quantity, besides $Y_1(\omega)$ of Eq.~\eqn{uno-wrong}, that we 
shall study to evaluate how wrong the replacement of $\bd{A}(t)$ by 
$\bd{A}_\text{ext}(t)$ in the minimal coupling scheme may be.

\begin{figure}
  \centerline{\includegraphics[width=0.5\textwidth]{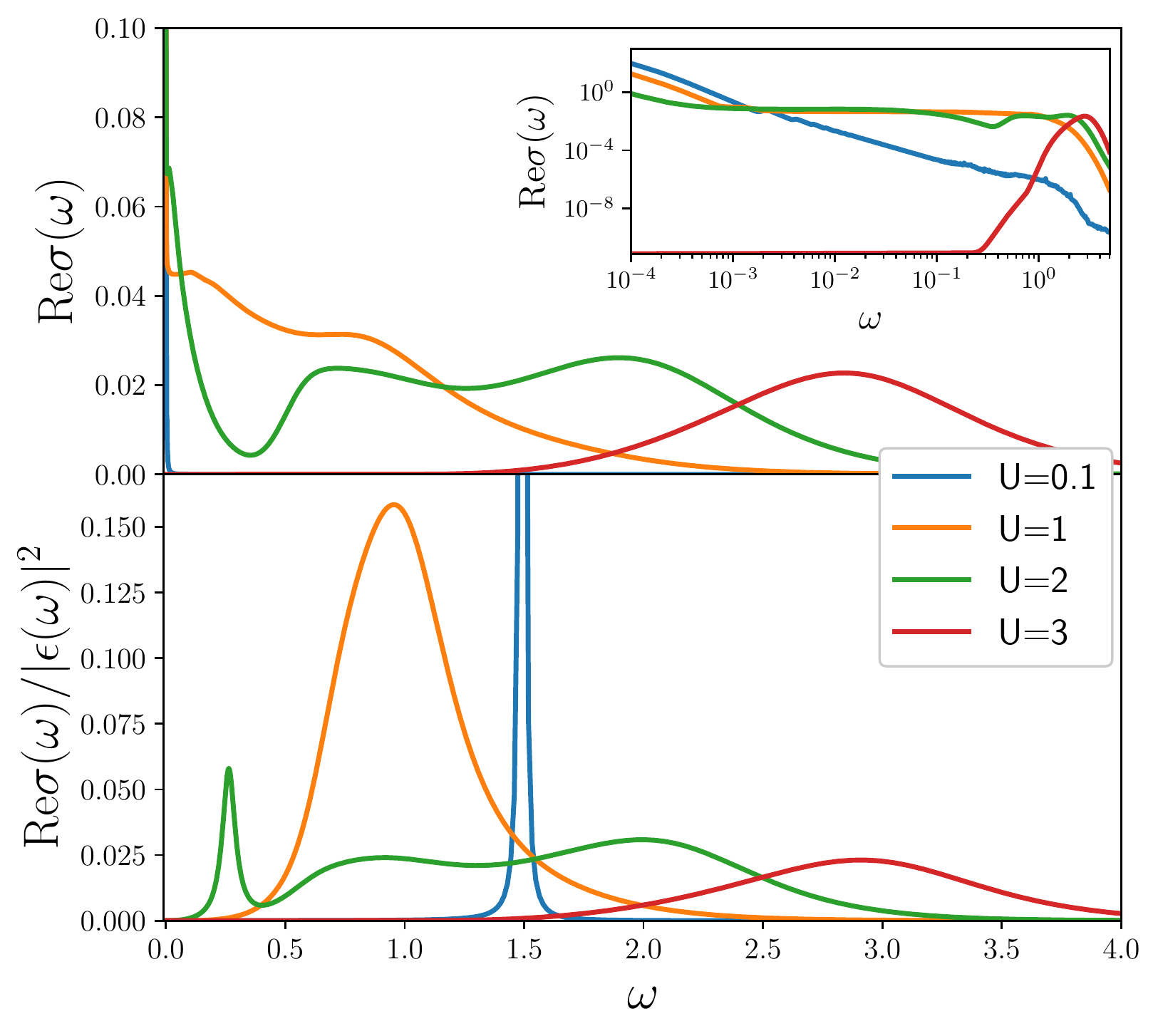}}
  \caption{(Color online)
    Top panel: absorption spectrum from the internal field as obtained
    from the optical conductivity. Inset: low energy behavior in
    logarithmic scale.
    Bottom panel: absorption spectrum from the external field. See main
    text for the defintion}
  \label{absorption}
\end{figure}
\section{Results}
We calculate at zero temperature the optical
properties of the Hubbard Hamiltonian \eqn{Ham-Hubb-1} at half-filling
by means of DMFT \cite{DMFT}, using numerical renormalisation group
(NRG) as impurity solver \cite{Hewson1993,Bulla-1998,NRG-DMFT-RMP2008}.
Specifically, we calculate the single-particle Green's function, 
through which we obtain the local single-particle spectral function, $A(\omega)$, and the uniform current-current response function \cite{DMFT,Rozenberg1995PRL,Tomczak2009PRB,Arsenault2013PRB,Zitko2015PRB}, which, in turn, allows computing the optical conductivity and thus the 
dielectric constant. 
In what follows we shall use as units of measurement the
half-bandwidth $8\mathsf{t}=1$, the lattice constant $a=1$,  the
electric charge $e=1$ and finally $\hbar=1$.

\begin{figure}
\centerline{\includegraphics[width=0.5\textwidth]{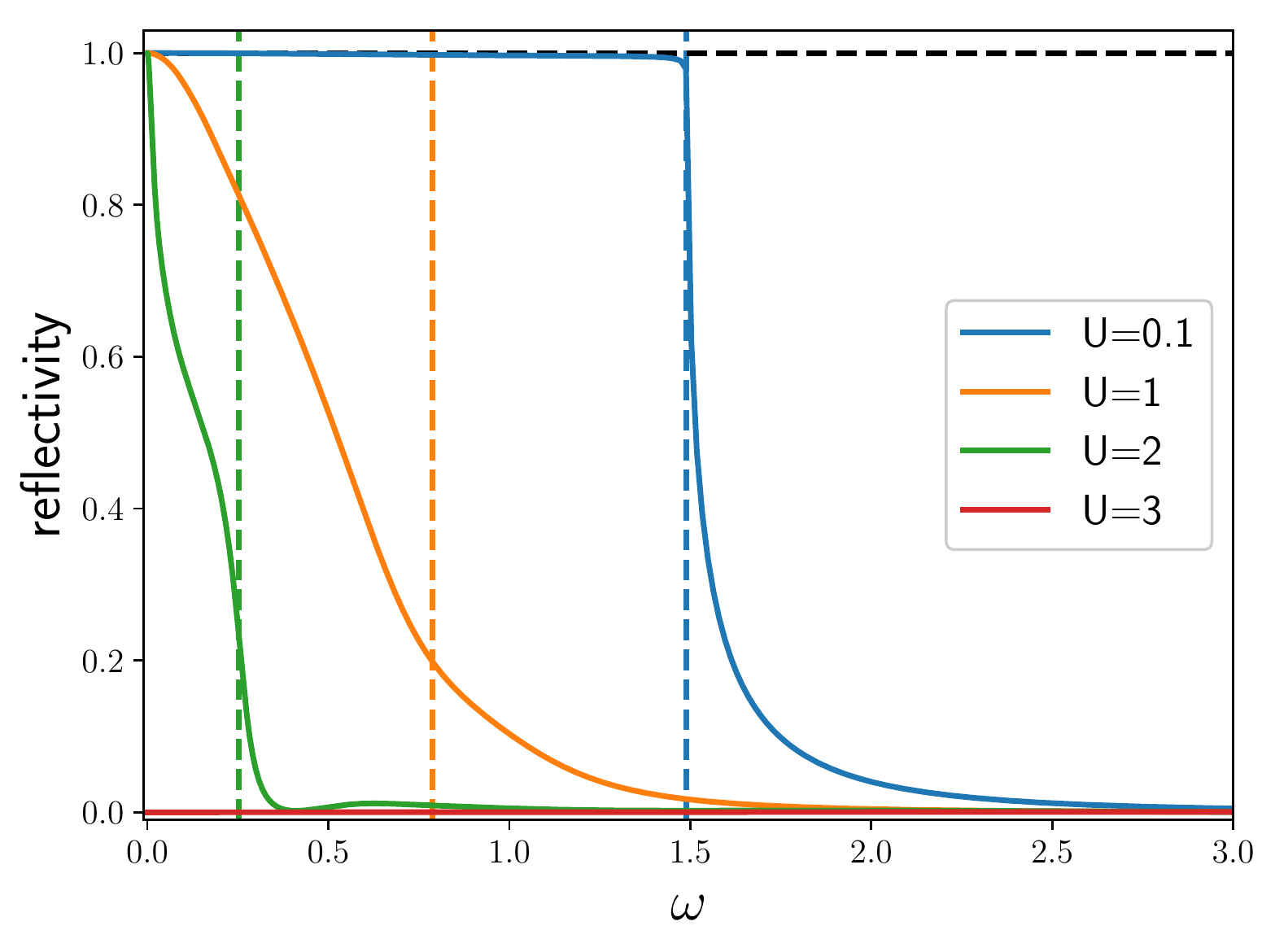}}
\caption{(Color online) Reflectivity as function of frequency. Data
  are for the same values of $U$ of previous figures.
  The vertical lines indicate the positions of the intra-band
  plasmons. 
  Note that the $U=3$ Mott
  insulator case is nearly vanishing. }
\label{reflectivity}
\end{figure}

To fix ideas, we show in Fig.~\ref{LDOS} the evolution of the local single-particle spectral 
functions $A(\omega)$ with increasing $U$ from the weakly correlated metal, $U=0.1$, 
up to the Mott insulator, $U=3$.
We note that for the intermediate
interaction strength ($U=2$) coherent quasiparticles narrowly peaked
at the chemical potential $\omega=0$ coexist with the lower
and upper forming Hubbard sidebands, centred at $\omega \sim \pm U/2$, respectively.

We now discuss the optical properties of the model from the weak
coupling metal to the Mott insulator~\cite{Rozenberg1995PRL,DMFT,Limelette2003S,Comanac2008NP}. 
According to Eq.~\eqn{due}, the absorption spectrum from the internal field is 
the real part of the optical conductivity, $\text{Re}\,\sigma(\omega)$, 
while that from the external field is instead
$\text{Re}\,\sigma(\omega)/|\ep(\omega)|^2$, shown, respectively, in the top
and bottom panels of Fig.~\ref{absorption}.

Looking at the top panel of Fig.~\ref{absorption}, we observe that the
optical conductivity of the weakly correlated metal at $U=0.1$
just shows a very narrow Drude peak.
This peak broadens upon increasing the interaction strength $U$.
Two additional absorption peaks emerge, which are most visible for
$U=2$:
an intermediate one involving an excitation from/to the quasiparticle peak to/from the Hubbard bands, 
and a high-energy peak corresponding to an excitation between the two
Hubbard bands. The latter is the only one that survives in the Mott
insulator at $U=3$.

\begin{figure}
\centerline{\includegraphics[width=0.5\textwidth]{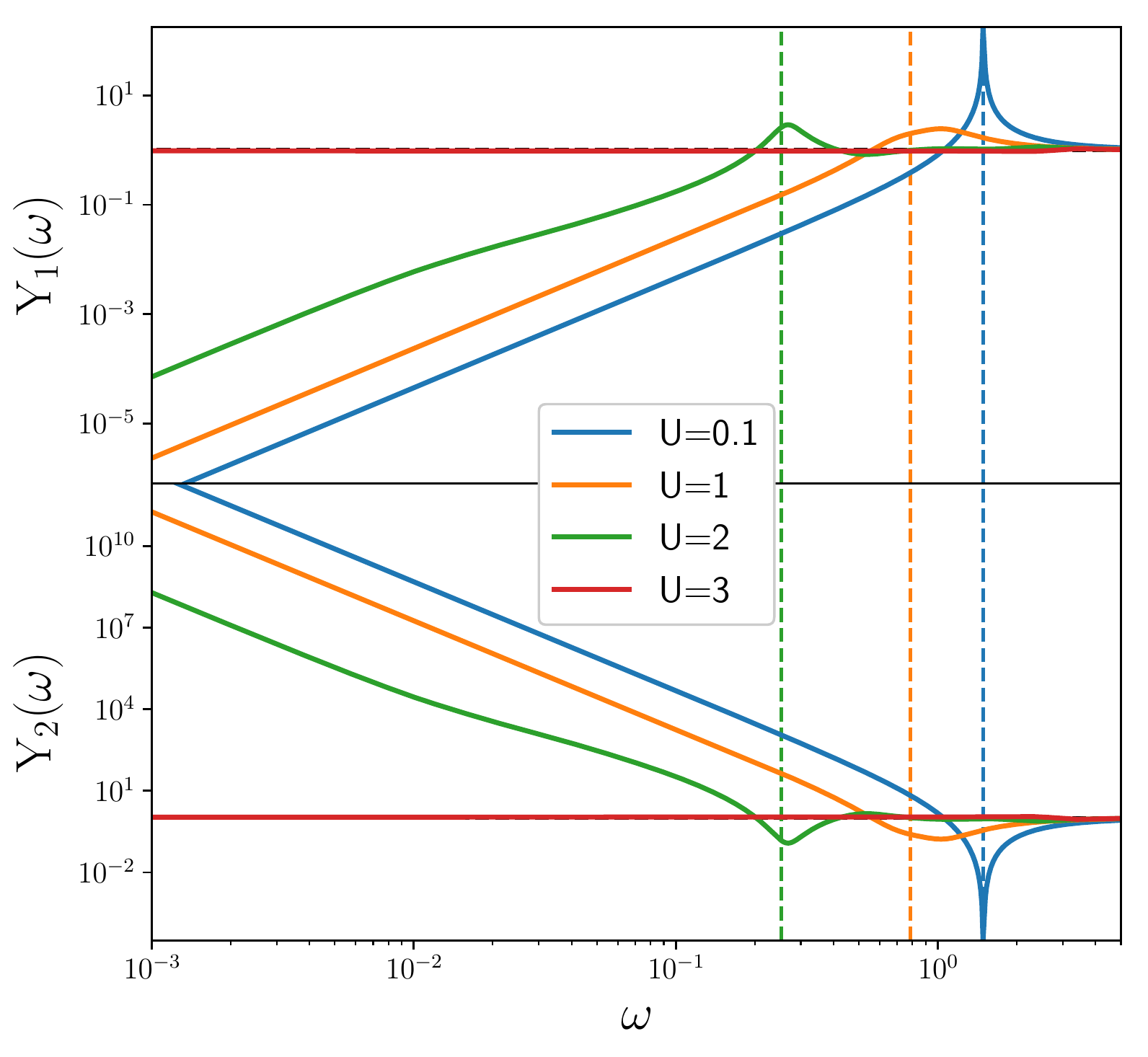}}
\caption{(Color online)
  The behavior of the functions $Y_1(\omega)$ (top panel),
  Eq.~\eqn{uno-wrong},  and $Y_2(\omega)$ (bottom panel), 
  Eq.~\eqn{Delta T vs.}. Data are for the same values of $U$ as in
  previous figures. The vertical lines indicate the roots of
  the real-part of the dielectric constant.}
\label{estimators}
\end{figure}

The absorption spectrum from the external field 
\beal
\fract{\text{Re}\,\sigma(\omega)}{\,\big|\ep(\omega)\big|^2\,}
= -\fract{\omega}{4\pi}\;\text{Im}\bigg(\fract{1}{\epsilon(\omega)}\bigg)\,,
\label{Im 1/ep}
\eal
is presented in the bottom panel of Fig.~\ref{absorption}.
This quantity is rather different from the internal field absorption
spectrum, being dominated by the plasmon modes, i.e., the peaks of
$\text{Im}(-1/\epsilon(\omega))$.
At weak coupling, $U=0.1$, there is just a single and very sharp
intra-band plasmon.
The plasmon peak shifts to lower frequencies upon increasing $U$. 
Meanwhile, additional inter-band, i.e., involving the Hubbard sidebands, broad plasma modes emerge,   
see the intermediate coupling case at $U=2$.
In Fig.~\ref{reflectivity} we show the corresponding reflectivity,
where the plasma edges become clearly visible.

We can now return to our original aim, and try to quantify through the
behavior of the quantities 
$Y_1(\omega)$ in Eq.~\eqn{uno-wrong} and $Y_2(\omega)$ in  
Eq.~\eqn{Delta T vs.} the error generated by using the external
vector potential in place of the internal one within the minimal
coupling scheme.
We show the functions $Y_1(\omega)$ and $Y_2(\omega)$ in
Fig.~\ref{estimators}. 
From the behaviour of $Y_1(\omega)$, top panel of Fig.~\ref{estimators}, 
we conclude that, in the metal phase and for frequencies smaller than the intra-band plasmon modes, defined by the roots 
of $\text{Re}(\epsilon(\omega))$, 
the external field required to significantly reduce the 
expectation value of the hopping is orders of magnitude larger than
what is predicted by assuming that $\bd{A}(t)$ in the Hamiltonian
\eqn{Ham} can be replaced by the external field $\bd{A}_\ext(t)$.
Within that same assumption and in the same range of frequencies, the
temperature raise produced by the field would be huge compared to the
actual value, see bottom panel in Fig.~\ref{estimators}.
On the contrary, and not surprisingly, $\bd{A}(t)\simeq
\bd{A}_\ext(t)$ works well in the insulating phase at $U=3$.

In conclusion, we have shown that replacing in the minimal coupling scheme the 
internal vector potential, which is self-consistently determined by
the system charges, by the external vector potential may be quite
dangerous, in particular in a metal and when the frequency of light is
small compared with the intra-band plasma edge, which is where
screening effects are maximal. 
In correlated metals the precise value of such plasma edge, which originates from the itinerant carriers and is proportional 
to the square root of their contribution to the optical sum rule
\cite{Millis-PRB2005,Basov-pnictides-2009,Degiorgi_2011}, is material
dependent \cite{Plasmon-cuprates,DegiorgiPRB1994,Basov-pnictides-2009,RevModPhys.83.471}
and typically ranges from mid to near infrared. This in turn implies
that in common pump-probe experiments the internal field $\bd{A}(t)$
is rather different from the external one $\bd{A}_\ext(t)$, hence replacing 
the former by the latter in model calculations is simply incorrect. 

We end mentioning that mixing up the response to the internal field with 
that to the external one is a mistake that tends to recur. It was, e.g., 
at the origin of early claims that the conductance of Luttinger
liquids is renormalised by interaction; a wrong statement corrected in
\cite{Finkelstein-PRB1996} by similar arguments as ours.

We acknowledge support by the European Research Council (ERC) under H2020 Advanced Grant No. 692670 ``FIRSTORM''.    
 

\bibliographystyle{apsrev}

\bibliography{mybiblio}

\begin{thebibliography}{47}
\bibitem[{irr()}]{irreducible}
\bibinfo{note}{For instance, the Feynman diagram expansion of the transverse
  response lacks diagrams that are reducible by cutting an interaction line,
  and which are singular in the long wavelength limit. The same diagrams must
  not be taken into account also in the case one studies directly the response
  to the internal field, either transverse or longitudinal. In both cases,
  restricting the perturbation series just to the diagrams irreducible to
  cutting an interaction line, one would not find qualitative differences
  between a long-range interaction and a short-range one.}
\end{thebibliography}

\end{document}